# Unified mobile public health care system (UMPHCS) for underdeveloped countries

Shadman Sakib, Rakibul Haq and Tariq Wazed
Department of Naval Architecture & Marine Engineering
Bangladesh University of Engineering & Technology
Dhaka, Bangladesh
rnd.in2gravity@gmail.com

*Abstract*—In this paper, we have proposed a new smartphone based system for health care, monitoring and diagnosis, which is specially designed to efficiently increase the public health care system in the distant, rural, unreached areas of the underdeveloped and developing countries. In this all-in-one system, we have digitized the health monitoring and diagnostic devices in a way so that each device works as a minimum 'plug and play' sensor module of the total system, reducing the cost radically. Besides, the easy-to-use smartphone application for operating the whole system reduces the necessity of skilled and trained manpower, making it a perfect toolbox for the government health workers in the unreached rural areas.

## I. INTRODUCTION

Health care equipment is today's one of the most rapidly growing engineering sectors. We can see a lot of research and improvements in this field. Like others, these equipments are now being digitized and computerized to improve the accuracy and bring ease to operation. We all are familiar with handy, small digital blood-pressure meter, blood sugar meter, etc. In spite of their familiarity, they are not so cheap. As a result, they remain out of the poor people's reach. In the underdeveloped and developing countries, setting up a public health care system of a minimum quality in the rural areas is still a huge challenge for the government due to the huge expense and scarcity of skilled manpower. In developing countries like Bangladesh, government hospitals in the rural areas do not have the minimum medical equipment. Even there is a serious scarcity of qualified doctors and technicians. Good doctors and specialists are not willing to go to the underdeveloped areas. Villagers' only hope is the government health workers who are a little educated and qualified. Besides, they do not have the necessary equipment for health checkup and diagnosis. These areas demand a cheap, but easy system for health monitoring, diagnosis and storing the results in a database so that government policy makers, researchers and specialists or doctors can monitor them from distant places.

Every digital health check-up device has a processing unit, a display unit and a sensor module. Now-a-days, smartphones have become very powerful and they come with high computational power and visual display units. So, why not use them as a medical device? They have the common part of every digital device- the CPU and the graphics unit. All we need is just a sensor module interfaced to the smartphone to convert it into a medical device. And just by changing the sensor module, it can become another device. So, the overall cost will be very little compared to buying those devices individually. Some researchers have worked to make smartphone based health check-up devices. But all of them are made to do just a certain test like eye power or blood pressure. Another school has developed a personal cloud based ubiquitous health monitoring system. These are based on the Wireless Body Area Sensor Networks (WBASN) technology [1], This technology is for personal use as one unit is needed for each person and certainly not suitable for the underdeveloped areas. We need a cheap and public system for these areas where people of a certain community can share one device. The device on which this paper is written has the flexibility to attach with various types of sensors to conduct different tests. Moreover, all of the data processing and calculations will be carried out on the phone. An assisting sensor hub will just collect the sensors' information and send it to the phone using serial protocol. We do not need to alter the hub and the phone for different tests; just the sensor module. And as the module will do nothing but just 'sensing', it will be very simple and of very low cost.

After the checkups, the results will be saved in a local database for each person and it is also possible to upload it to a server for the specialist doctors who will review them regularly. This database will also help us to determine the overall and individual health condition of the people of certain underdeveloped regions and eventually the whole country.

## II. RELATED WORKS

Previously, all works done in the mobile health care system were mainly to develop a personal mobile health monitoring system [2] [3] [4] and a few low cost hardware-smartphone interfacing projects for certain health tests [5].

Researchers of the first stream mainly focused on developing a smartphone based system where the smartphone gets data via Bluetooth from different sensors locating on different places of the user's body, each measuring different health conditions like pulse, temperature, blood pressure etc. It is a system for personal use. These systems continuously update the user's health condition in a server making it possible to remotely detect any unusual or sudden health problems like heart attack or any other accident instantly and taking actions. This system is probably the future of our health care system but not applicable for the underdeveloped countries where people do not afford to have these types of expensive personal systems. In this paper, we have tried to develop such a system. We too used the smartphone to process the data received from the sensors but have also implemented the flexibility to make it possible to use each single unit for public health care instead of using it only personally. In doing so, it will be possible for the government health workers to easily use this system for health monitoring and diagnosis of all the people of a certain area and it is completely unnecessary for the subjected people to possess a smartphone. This system provides the facility to store individual health data in a server, like the present WBASN technology based systems. So, the government will directly get the health data of the inhabitants of the unreached area instantly which is a great problem for the underdeveloped countries. Moreover, the system uses a physical interface between the external circuit and the phone, eliminating the necessity of the expensive Bluetooth modules. Though, because of the fact that some phones do not have the USB host feature which is necessary to connect to the external circuit, we have also kept an alternate Bluetooth connection method in the system. Again sharing same circuitry and even in few different cases, the sensor made the system ultra-cheap and so, a perfect option for the underdeveloped countries.

In our research, we tried to integrate many diagnosis devices under one universal platform to build a ubiquitous health care system and also keeping the flexibility for public use. It is mainly designed in a way so that one unit can be used for a whole community rather for a person. But it has that flexibility too. In a sense, we implemented a WBASN type technology for public healthcare system which was never done before.

### III. SYSTEM OVERVIEW

Our proposed system, which we have named Unified Mobile Public Health Care System or UMPHCS, has four parts. They are:

*A. Sensor module:*

This module is used to detect various physical properties of the patient. Like temperature, blood pressure, weight, blood sugar, etc. This module merely creates an electrical signal depending on the target properties. Like a variable voltage in case of temperature sensing. In this particular case, the sensor module is just an LM35 temperature sensor, nothing else. Others are similarly small, consisting of the minimum circuitry and sensor arrangement to produce an electrical signal which depends on the target property, mostly an analog voltage. Even tests which need same sensors, will share one sensor to reduce the cost. They will consist of two sub-modules, among which one will be the sensor part and the other is the arrangement to connect it with the body or test subject.

*B. Sensor hub:*

This is a microcontroller-based small circuit. It is connected to the smartphone via serial connection and has a port for attaching different sensor modules. This port will have supply voltage pin, ground and Analog to Digital Conversion (ADC) channel of the microcontroller. In our project we did not need any other pins. But in the future, this might be enhanced and in fact, our aim is to bring as much medical equipment as possible under this system, some of which might need some more pins. So, the pin configuration of the hub can be altered. Different modules will only connect with the pins they need. The job of the sensor hub is always to collect the sensor data (like ADC value) and send it to the phone. Even it will not know which sensor it is attached to. Again, this can be enhanced on the basis of necessity.

*C. Smartphone:*

The whole system is smartphone based. So, clearly this is the most important part of the system. The operating software of this UMPHCS will be installed into a smartphone. Every detail of the sensor modules will be in that software. After attaching a certain sensor module to the hub, the user will select the corresponding diagnosis option from the software's menu. Then the software will run the code block for that diagnosis. The hub will provide it with the sensor value needed to determine the attribute of the physical property being tested. After displaying on the screen, the result of each diagnosis of a person will be saved instantly in a local database on the phone. For abnormal result, the software will give some suggestions too. Later the user will upload it to the master online database. This will also be done by the same software.

*D. Online Database:*

The online database will be for the specialists who will examine these individual data for any anomaly, abnormal pattern in one's medical history and any sign of upcoming epidemic. As an example, gradual decrease of someone's weight is a symptom of many diseases like AIDS. Again, if this happens to most of the inhabitants of a region, we might be a sign of an upcoming epidemic and if we can predict this, we can take necessary steps to prevent it or lessen the harm. So, definitely this individual database will be a great tool for the health researchers and the health policy makers of the government. It will also help the statistical researchers too.

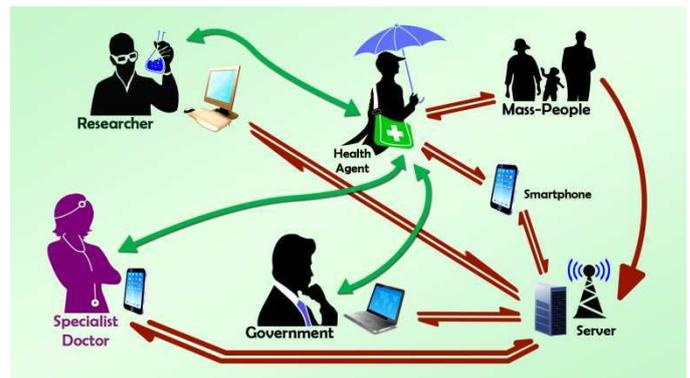

Fig. 1 System overview of UMPHCS

### IV. PROTOTYPE

For testing the efficiency and accuracy of the UMPHCS, we have built a prototype able to perform eight different physiological tests. They are-

1) Body temperature.
2) Blood pressure.

3) Heart rate.
4) Weight.
5) Eye power.
6) Hearing test.
7) Height.

All of them use the same sensor hub to connect to the phone. Each sensor is designed and fabricated by ourselves. Some modules like eye power measuring one are made using unique concept. Details of the sensor modules, sensor hub, host smartphone and the operating software will be discussed separately in this section.

*A. Sensor modules:*

We have used four sensor modules in total, for six different tests. Besides these, we have also used the smartphone's own camera and speaker to measure the height and perform the hearing test, respectively. It is done to show that the present smartphones have some resources, like accelerometer, camera, speaker, GPS which can be used for various tests, which will reduce the cost even more while increasing the operational simplicity. Detail description of the modules is given bellow:

*1) Temperature measurement module:* This is made with a low-price very common temperature sensor- LM35. It is the simplest module of all. The $+V_s$ and GND terminal respectively get connected to the hub's voltage source and ground pin. And the $V_{out}$ pin connects with the ADC channel. The voltage on the $V_{out}$ pin depends on the temperature of the sensor. The module is designed in a way so that the LM35 remain on the tip of a thermometer-like stick. So, it can be used as a traditional thermometer and by analyzing the voltage level, it is possible to detect the body temperature very easily.

*2) Blood pressure and Heart rate measurement module:* This module is just a simple arm cuff of the traditional blood pressure measuring device without the sphygmomanometer. This submodule gets attached to a pressure sensor by the pipe-end where normally the sphygmomanometer gets attached. We have used MPVX5050GP for measuring the air pressure.

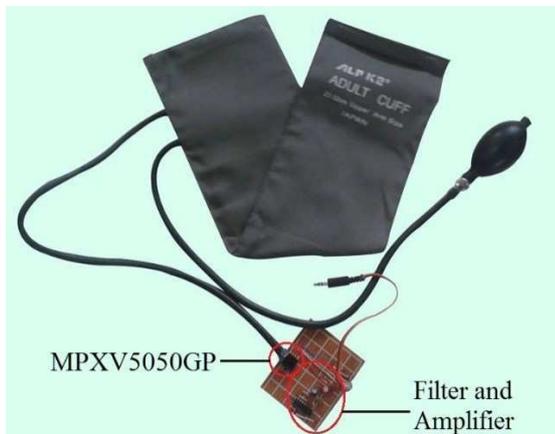

Fig. 2 B.P. and H.R. measurement module

*3) Weight measuring arrangement:* It is a simple module based on load sensor. The output of the load sensor is an analog voltage which is proportional to the applied load or weight. But the load sensor's output signal is too small to detect with the microcontroller's ADC pin. Its highest value is 2 mV/V. So, we needed an instrumentational amplifier to amplify the output signal to the convenient region for the microcontroller. For this purpose, we have used an AD623 instrumentational amplifier.

*4) Eye power measurement module:* This module can determine the power of glasses a myopic or hyperopic person needs. It is made with just two lenses and a sliding variable resistor or pot. The lenses are attached inside a rectangular hollow pipe. One lens is fixed in a position near one end, which is the eyepiece. Another lens is not fixed. Its position can be altered linearly to adjust the distance between the lenses. The sliding pot will be attached to this lens. The user will look through the eyepiece to a distant object or writing and slide the pot's knob to adjust the position of the lens in a certain position where he/she can see clearly. As the distance between the lens changes, the equivalent focal length changes too. We can find it from the equation below:

$$F' = \frac{f_1 \times f_2}{f_1 + f_2 - D} \quad (1)$$

Where, $F'$ = Equivalent focal length of the used lenses,
$f_1$ = focal length of the first lens,
$f_2$ = focal length of the first lens
And $D$ = the distance between the lenses.

As the resistance of the pot changes linearly, it actually changes linearly with $D$.
So, we can say, $D = C \times V_{ADC}$ (2)
Where, $C$ is a constant.
Again, the virtual position of the equivalent lens,

$$\alpha = \frac{F' \times D}{f_2} \quad (3)$$

But, when a person wears glasses, the glasses remain at a fixed distance from the eye, say, $L$. So, we have to find the equivalent lens of this combination at distance $L$. If this equivalent lens is $F$ and the distance between the user's eye and the eyepiece is $l$, then-

$$L = \frac{F \times (l + \alpha)}{F'}$$

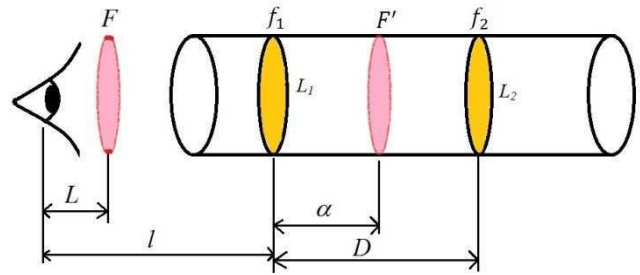

Fig.3 Arrangement of the eye power measurement module

Or, $$F = \frac{F' \times L}{(l + \alpha)} \quad (4)$$

And, $$P = \frac{1}{F} \quad (5)$$

Combining equation (1), (2), (3), (4) and (5), we get-

$$P = \frac{(l \times P_1) + (l \times P_2) + (D \times P_2) - (l \times D \times P_1 \times P_2)}{L} \quad (6)$$

Where, $P_1$ and $P_2$ are the power of $L_1$ and $L_2$ respectively.

After some inspections, we have found that the glasses of the spectacles usually remain 1.5 cm in front of the eye lens. In our device, we have placed the first lens 3 cm away from the eye. So, putting these values of $L$ and $l$, we get-

$$P = \frac{(D \times P_2) - (0.03 \times P_2) - (0.03 \times P_1) + (0.03 \times D \times P_1 \times P_2)}{0.015} \quad (7)$$

After some calculations, we have used lenses of power +5.00D and -2.00D as $L_1$ and $L_2$ respectively. This configuration gives us a range of measuring the necessary glass power from -1.30D to 17.50D perfectly. Initially we only develop this module for hyperopic patient. So, it does not cover the myopia side. It is possible to build one easily.

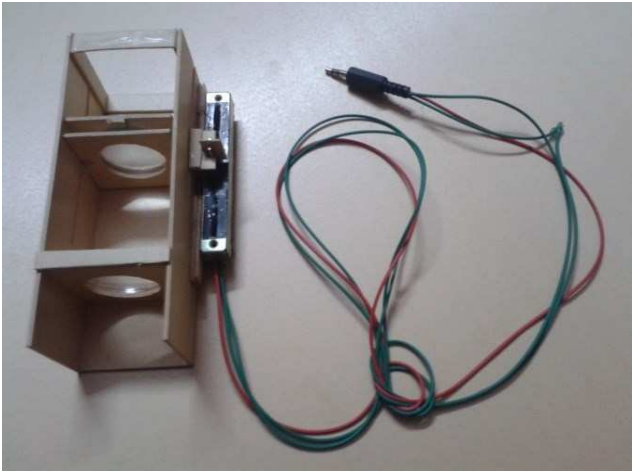

Fig. 4 Eye power measurement module with upperside open to show the internal arrangements and mechanism.

TABLE I.  RANGE OF EQUIVALENT LENS'S POWER FOR SELECTED TWO COMBINATION

| Distance (cm) | P for +5D and -2D (for hypermetromia patients) | P for -5D and +2D (for myopia patients) |
|---|---|---|
| 1.5 | -1.3 | 0.7 |
| 2 | 0.266667 | -1.06667 |
| 2.5 | 1.833333 | -2.83333 |
| 3 | 3.4 | -4.6 |
| 3.5 | 4.966667 | -6.36667 |
| 4 | 6.533333 | -8.13333 |
| 4.5 | 8.1 | -9.9 |
| 5 | 9.666667 | -11.6667 |
| 5.5 | 11.23333 | -13.4333 |
| 6 | 12.8 | -15.2 |
| 6.5 | 14.36667 | -16.9667 |
| 7 | 15.93333 | -18.7333 |
| 7.5 | 17.5 | -20.5 |

*5) Hearing test:* This test is done by creating a sound wave and sending it through a simple headphone to the patient's ears. The frequency of the generated sound wave gradually increases as well as the amplitude from -5dB to 80dB in each frequency. The frequency starts from 250Hz and ends at 8000Hz. Whenever the amplitude reaches the upper level, the frequency increases and the amplitude again starts from -5dB. The user waits till he/she hears a tone. Whenever he/she gets it, the operator will move onto the next frequency. After going through each frequency needed for a complete audiogram, the hearing test will stop and the all heared amplitudes in those frequencies will be calculated to show hearing limit of the user with the audiogram graph.

*6) Height measurement:* Height is measured by using the camera of the smartphone. Though this measurement is better done by hand, we included this tool too. For height measurement, the target person stands straight with an accurate ruler in his/her hand aligning with his/her body in front of the phone's camera. Then the operator selects height measurement option from the menu. This opens the camera to take a photo. After taking a photo, the software tells the operator to point out the both end points of the ruler as well as the tip of the head and foot of the target person. The software then measures the pixels between the end points of the ruler and the person. Then it finds the height of the person from the following simple equation-

$$Height = \frac{N_B}{N_R} \times length\ of\ the\ ruler$$

Where, $N_B$ = Number of pixels between persons head and feet. And $N_R$ = Number of pixels between the ends of the ruler.

*B. Sensor hub:*

The sensor hub is based on an AVR microcontroller. The hub has an input port for the sensors consisting of three terminals- $V_{cc}$, GND and ADC as mentioned before. This hub just converts the analog voltage of its input channel into digital number and sends that value to the phone via serial Rx-Tx. An USB to serial converter chip was needed to interface the microcontroller with the phone or tablet. For rapid prototyping, we have used "Arduino nano", a renowned cheap prototyping board [8] to build the hub. We have used the V3.0 version. It is based on an ATmega328 microcontroller, which runs at 16MHz clock speed and has an FTDI FT232RL chip for USB to serial conversion. The hub always waits for a command in the serial line. Meanwhile it does nothing. After getting the incoming command it takes a sample from its ADC channel, convert it into a number and send in back to the phone via serial line. This whole program took only 2.35 KB memory where maximum available space in Arduino Nano V3.0 is 30.72 KB. We have used the traditional stereo audio jack as an input port. Because of its design, while plugging in or out, the sensor modules may have a short circuit. So, we have included a safety cutoff switch and an indicator LED for module changing. In mass production this board will be replaced by a simpler circuit board based on a more simple microcontroller to reduce the cost, as we do not need anything but the ADC and serial communication in the sensor hub.

As most of the phones do not have the USB host feature, we have also built a Bluetooth based sensor hub, which is a little more expensive because of the Bluetooth module. This Bluetooth sensor hub also uses Arduino nano. But it has an extra rechargeable battery pack of 7.5V and a cheap HC-05 Bluetooth module. The power consumption is greater in the Bluetooth sensor hub. It's current consumption varies from 90-100 mA depending on attached modules. So, the maximum power consumption is around 0.75 Watt.

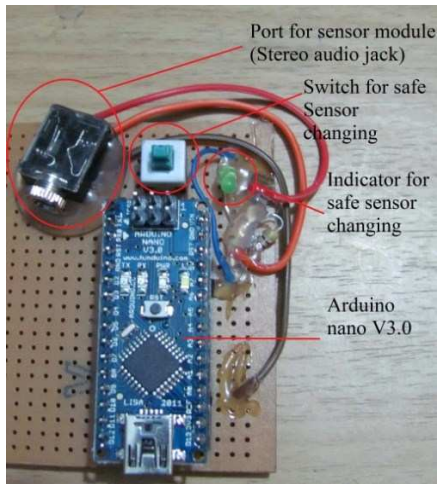

Fig. 5 Wired Sensor hub

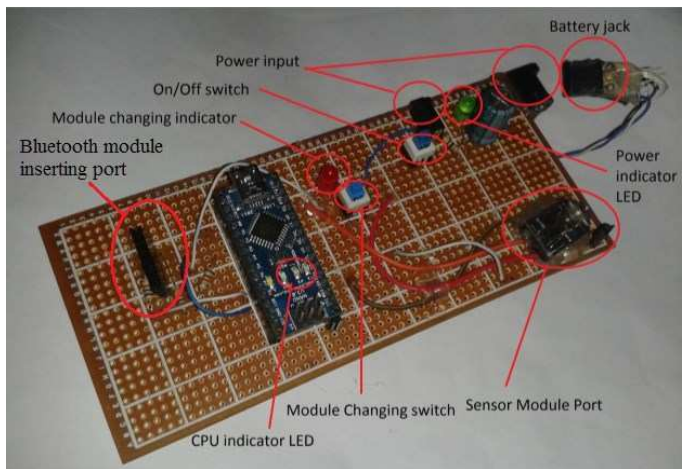

Fig. 6 Bluetooth Sensor hub

### C. Smartphone and the software:

We have used a Symphony T7i android operated tab to test the prototype using wired sensor hub. It runs with the Android version 4.0.3, 1.2GHz dual core processor and 1GB of RAM. The software uses the serial port. So, it needs the USB host feature and a minimum API level 10. On the other hand, the Bluetooth sensor hub does not have this limitation. It can connect to any mobile phone or tab having Bluetooth. We have also built a software for windows phone to test the Bluetooth connection. We need WP8 or higher OS to use the Bluetooth feature. In our research, we have used Nokia Lumia 525 with WP8. It also has a 1 GHz dual core processor and 1 GB RAM.

The algorithm of the software side of the smartphone is simple. Apart from the Blood pressure measurement, for all other cases, the phone sends a character as command to the sensor hub when "Calculate" button is tapped and waits for incoming data which is the ADC value. Then, inside the software, those values are converted into temperature, eye power and weight using specific equations. In case of blood pressure, the software sends command in a timely basis as we need a data stream rather than a single value to calculate S.P. and D.P. The time interval is 10 mili seconds. To get the S.P. and D.P. we need to process the pressure data to get the oscillation waveform. This can be done by digital signal processiong inside the phone, which we have used. But using a bandpass filter circuit increases the accuracy. So, we have also kept an option there. The BP and HR module has a built-in bandpass filter. But using this filter requires an extra ADC channel. In that case, the sensor hub will sample the filtered data to get the oscillatoion waveform directly and find the SP and DP from it. Many research have been done to find out the SP and DP from the oscillatoion waveform. [14][15][16] We have taken the points of 50% and 70% of the maximum oscilation of the OW as SP and DP respectively for optimum result.

The software demands the operator to select a specific person before starting tests by either typing a new name or selecting from previous database and saves the results in a local SQLite database which can be viewed anytime later.

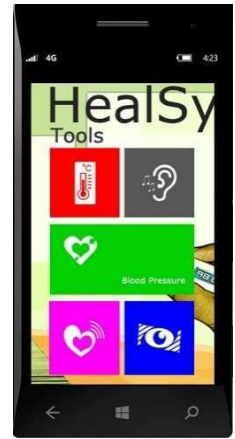

Fig.7 Developed software, HealSys. (For windows phone 8 platform).

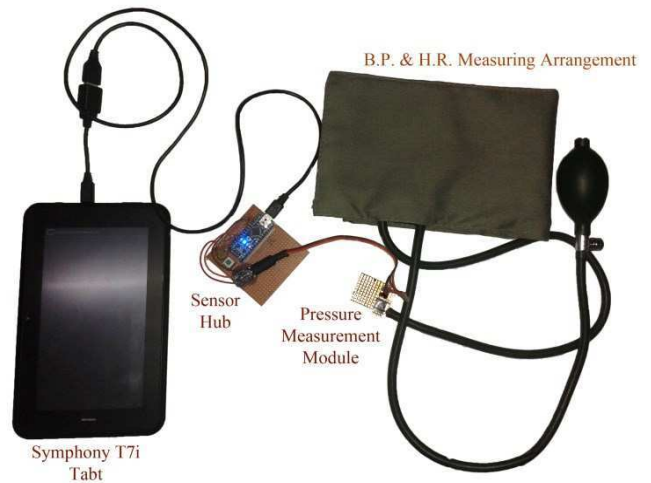

Fig. 8 The B.P. and H.R. measuring arrangement is connected to the USB port of an Android running Tab through the pressure measurement module and the sensor hub.

## V.　RESULTS

We have tested our prototype by measuring data which were already known. The known values were found by taking the measurements in the trusted traditional way. Then, for some cases, we needed to calibrate the equations for optimum result.

But after the calibration, we got a fair accuracy for most of the modules. Details are shown in Table II.

TABLE II. ACCURACY OF THE MODULES

| Digital device | Accuracy |
|---|---|
| Thermometer | Calibrated to give the same result as the reference thermometer did. |
| Blood pressure & heart rate meter | Fluctuates upto 8 mmHg comaparing to a renowned digital BP machine. |
| Weight meter | Gives the same result as other digital weight meters do. |
| Eye power measuring device | Calibrated for better result. But still fluctuates person to person. |
| Ear power measuring device | Accurate. |
| Height measuring device | Fluctuates upto 4% depending on the edge selection accuracy. |

As our main concern was to make this system a cheap one, we did a survey of present market price of the digital medical test equipments which we have included in our prototype.

TABLE III. MARKET CONDITION OF THE DIGITAL HEALTH TEST DEVICES WE HAVE INCLUDED IN THE PROTOTYPE

| Digital device | Market price |
|---|---|
| Thermometer | 100 BDT |
| Blood pressure & heart rate meter | 2500 BDT |
| Weight meter | 3000 BDT |
| Eye power measuring device | N/A |
| Ear power measuring device | N/A |
| Height measuring device | N/A |
| Total | 5950 BDT |
|  | = 76.28 USD |

TABLE IV. DETAIL EXPENDITURE FOR BUILDING THE PROTOTYPE

| Equipments | Market price |
|---|---|
| LM35 | 50 BDT |
| MPXV5050GP | 1200 BDT |
| Load sensor and instrumentational Amplifier | 1200 BDT |
| Cuff | 350 BDT |
| Lenses | 300 BDT |
| Arduino nano | 700 BDT |
| Accessories | 500 BDT |
| Total | 4300 BDT |
|  | = 55.13 USD |

## VI. CONCLUSION

From Table II it is seen that the height measurement and BP module has a little accuracy problems. Accuracy of the height measurement can be improved by using stylus instead of finger to select the edge. The BP module has limitations, but all digital BP machine has accuracy problem in this range. Moreover, our BP module aquires data in the same way other digital BP machines do. So, better circuitry and DSP techniques for them can be implemented here too. This accuracy will be more satisfactory when industrial production will take place. Moreover, mass production will also cut down the price significantly.

## VII. FUTURE WORK

We are currently working on integrating Glucometer, Peak flow meter and Electrocardiogram (ECG) modules into this system. We will also keep a callibration option in our next version of the mobile software which is very much needed.